\input{aipcheck}

\documentclass[
    ,final            
  ]
  {aipproc}

\layoutstyle{6x9}

\def\lsim{\raise0.3ex\hbox{$\;<$\kern-0.75em\raise-1.1ex\hbox{$\sim\;$}}}
\def\gsim{\raise0.3ex\hbox{$\;>$\kern-0.75em\raise-1.1ex\hbox{$\sim\;$}}}

\def    \bea           	{\begin{eqnarray}}
\def    \eea           	{\end{eqnarray}}

\begin{document}

\title{Phenomenology of a New Supersymmetric Standard Model: The $\mu\nu$SSM}


\author{Carlos Mu\~noz}{  
address={Departamento de F\'{\i}sica Te\'{o}rica UAM and Instituto de F\'{\i}sica Te\'{o}rica UAM/CSIC,\\ Universidad Aut\'{o}noma de Madrid (UAM),
Cantoblanco, E-28049 Madrid, Spain}
}



\begin{abstract}
The $\mu\nu$SSM solves the $\mu$ problem of the MSSM and explains the origin of neutrino masses by simply using right-handed neutrino superfields. The solution implies the breaking of R-parity. The properties and phenomenology of the model are briefly reviewed.



%
\end{abstract}

\maketitle



The LHC
will finally start operations at the end of this year.
It will be able to answer not only the crucial question about
the origin of the mass, but also to clarify whether or not supersymmetry (SUSY) exists.
SUSY not only solves several important theoretical problems of the Standard Model, but also has spectacular experimental implications.

However, it is fair to say that SUSY has also its own theoretical problems, and, in particular, a crucial one is the so-called
$\mu$-problem of the MSSM.
As it is well known, the NMSSM,
provides 
a
solution via the introduction of an extra singlet
superfield $\hat S$.


On the other hand, neutrino experiments have confirmed during the last years that
neutrinos are massive.
Thus all models must be modified in order to reproduce this result.
The BRpV 
has been proposed in this context.
There, bilinear terms breaking R-parity of the type, $\mu_i \hat L_i \hat H_2$, 
are added to the MSSM. These induce
neutrino masses 
through the mixing with the neutralinos (actually one mass at tree level and
the other two at one loop)
withouth including
right-handed neutrinos in the model, unlike the MSSM.
However, the $\mu$-problem is augmented with the three new bilinear terms.

The ``$\mu$ from $\nu$'' Supersymmetric Standard Model
($\mu\nu$SSM) 
was proposed in \cite{MuNuSSM,MuNuSSM0}
as an alternative to the MSSM, 
solving the
$\mu$-problem and explaining the origin of neutrino masses
by simply using right-handed neutrino superfields.

The superpotential of the $\mu\nu$SSM 
contains, in addition to the
Yukawas for quarks and charged leptons,
Yukawas for neutrinos
$\hat H_u\,  \hat L \, \hat \nu^c$, terms of the type
$\hat \nu^c \hat H_d\hat H_u$ producing an
{\bf effective  $\mu$ term} through right-handed sneutrino VEVs of order the
electroweak (EW) scale, $\mu\equiv
\lambda_i \langle \tilde \nu^c_i \rangle$,
and also terms of the type $\hat \nu^c \hat \nu^c \hat \nu^c$  
avoiding the existence of a Goldstone boson and contributing to generate
{\bf effective Majorana masses} ($\sim \kappa \langle \tilde \nu^c_i \rangle$):
\begin{eqnarray}
W & = &
\ \epsilon_{ab} \left(
Y_{u_{ij}} \, \hat H_u^b\, \hat Q^a_i \, \hat u_j^c +
Y_{d_{ij}} \, \hat H_d^a\, \hat Q^b_i \, \hat d_j^c +
Y_{e_{ij}} \, \hat H_d^a\, \hat L^b_i \, \hat e_j^c +
Y_{\nu_{ij}} \, \hat H_u^b\, \hat L^a_i \, \hat \nu^c_j 
\right)
\nonumber\\
& - &
\epsilon{_{ab}} \lambda_{i} \, \hat \nu^c_i\,\hat H_d^a \hat H_u^b
+
\frac{1}{3}
\kappa{_{ijk}} 
\hat \nu^c_i\hat \nu^c_j\hat \nu^c_k \ .
\label{superpotential}
\end{eqnarray}
Clearly, the above terms produce the {\bf explicit breaking of R-parity} 
(and lepton number)
in this model. The size of the breaking can be 
understood better if we realize that in the limit 
where neutrino Yukawa couplings $Y_{\nu}$ are vanishing, the 
$\hat \nu^c$ are 
ordinary singlet superfields like the $\hat S$ of the NMSSM, 
without any connection with neutrinos,
and
this model {\bf would coincide with the
NMSSM} 
(but with three singlets instead of one)
where R-parity is conserved .
Once we switch on the $Y_{\nu}$,
the 
$\hat \nu^c$ become right-handed neutrinos, and, as a consequence, R-parity
is broken. This breaking has to be small because we have an {\bf electroweak-scale seesaw}, implying that the values of $Y_{\nu}$ 
are no larger than $10^{-6}$ (like the electron Yukawa) to reproduce the neutrino masses ($\lsim 10^{-2}$ eV).

Actually, the breaking of R-parity
produces the mixing of neutralinos with
neutrinos, and as a consequence a generalized $10\times 10$ matrix of the EW seesaw type, 
\begin{equation}
{\mathcal{M}}=\left(\begin{array}{cc}
M & m\\
m^{T} & 0_{3\times3}\end{array}\right),
\label{matrizse}
\end{equation}
that gives rise at three level to three light eigenvalues corresponding to the neutrino masses \cite{MuNuSSM, MuNuSSM2}. Here $M$ is a $7\times 7$ matrix showing the mixing of neutralinos with right-handed neutrinos, and $m$ a $7\times 3$ matrix representing the mixing of neutralinos with left-handed neutrinos, and left- with right-handed neutrinos.
The entries of the matrix $M$ are of the order of the EW scale and much
larger than the ones of the matrix $m$ which turn out to be 
of the order 
of the Dirac masses for the neutrinos 
($Y_{\nu} \langle H_u^0 \rangle \lsim 10^{-4}$ GeV).
The latter can easily be understood, since
%
%
the entries of $m$ are proportional to $g\langle \tilde \nu_i \rangle$, 
$Y_{\nu}\langle \tilde \nu_i^c \rangle$ and
$Y_{\nu} \langle H_u^0 \rangle $. On the one hand, 
$ \langle \tilde \nu_i^c \rangle \sim \langle H_u^0 \rangle$. On the other hand,
because of the contribution of $Y_{\nu}\lsim 10^{-6}$
to the minimization conditions for the left-handed neutrinos,
the $\langle \tilde \nu_i \rangle$ turn out to be small, and one can check that they are 
no larger than $Y_{\nu} \langle H_u^0 \rangle$ 
\cite{MuNuSSM}.

Let us finally remark that having a (dynamical) EW seesaw avoids the introduction of {\it ad-hoc} high energy scales
in the model, as it occurs in the case of a GUT seesaw.

Notice that the neutrino Yukawas generate, through the $\langle \tilde \nu_j^c \rangle$
three efective bilinear terms,
$\mu_i \hat L_i \hat H_2$, with
$\mu_i\equiv Y_{\nu_{ij}} \langle \tilde \nu_j^c \rangle  \lsim 10^{-4}$ GeV. These
{\bf characterize the BRpV model}, as mentioned above.
{\bf The advantages} of the $\mu\nu$SSM (from our viewpoint) with respect to other popular models proposed in
the literature are now more clear. Concerning the $\mu$-problem, one solves it
without having to introduce
an extra singlet superfield as in the NMSSM. A special form of the
Kahler potential as in the Giudice-Masiero mechanism,
or specific
superpotential couplings to the hidden sector \cite{condensados,condensados2},
are not necessary either.

Using the eight minimization conditions for the scalar potential (which includes
the usual soft, $D$ and $F$ term contributions),
one can eliminate e.g. the soft masses $m_{H_d}$, $m_{H_u}$, $m_{\widetilde{L}_{i}}$, and $m_{\widetilde{\nu}_{i}^{c}}$
in favour of the VEVs of the Higgses and neutrinos.
We thus consider as independent {\bf parameters} of the neutral scalar sector \cite{MuNuSSM2}:
\begin{equation}
\lambda, \, \kappa,\, \tan\beta, \, \nu_1, \,  \nu_3, \nu^c, \, A_{\lambda}, \, A_{\kappa}, \, A_\nu\ ,
\label{freeparameters2}
\end{equation}
where
$\nu_i\equiv \langle \tilde \nu_i \rangle$, 
$\nu^c\equiv \langle \tilde \nu^c \rangle $, and 
we have assumed for simplicity that there is no intergenerational mixing
and that in general they have
the same values for the three families.

In the case of the neutrino parameters, 
following the discussion in \cite{MuNuSSM2,neutrinos}, it is sufficient 
with two generations with different VEVs and couplings in order to reproduce the experimental pattern. Thus we have chosen 
$Y_{\nu_1} \neq Y_{\nu_2}= Y_{\nu_3}$ and $\nu_1\neq \nu_2=\nu_3$.
Actually, we have checked that with
$Y_{\nu_2}= Y_{\nu_3} \approx 2 \; Y_{\nu_1} \sim 10^{-6} $ and
$\nu_2=\nu_3 \approx 2 \, \nu_1 \sim 10^{-4}$~GeV, the observed neutrino masses and mixing angles are reproduced (thus this result
can be obtained even with a diagonal neutrino Yukawa matrix
as pointed out in \cite{Ghosh:2008yh}). As explained in detail in \cite{neutrinos}, 
this is so easy to get due to the peculiar characteristics of this seesaw,
where the relevant scale is the EW one, and R-parity is broken involving not only 
the right-handed neutrinos in the seesaw but also the MSSM neutralinos.
In a sense, this gives an answer to the question why the mixing angles are so 
different in the quark and lepton sectors.
In \cite{neutrinos}, it was also shown that {\bf spontaneous CP violation} through complex VEVs is possible in the $\mu\nu$SSM at tree level.

The parameter space of the model was analyzed
in detail in \cite{MuNuSSM2}, studying the viable regions which avoid false minima and tachyons, as well as fulfill the Landau pole constraint. The structure of the 
{\bf mass matrices} and the
associated {\bf particle spectrum} was also computed.
The breaking of $R$-parity generates complicated mass matrices and
mass eigenstates, as we already saw above for the case of the neutralinos/neutrinos.
The charginos
mix with the charged leptons giving rise to a 5$\times$5 matrix.
Nevertheless, there will always be three light eigenvalues corresponding to the 
electron, muon and tau.
Concerning the scalar mass matrices,
the neutral Higgses are mixed with the sneutrinos,
and the charged Higgses with the charged sleptons,
and we are left with fifteen (eight CP-even and seven CP-odd) neutral states 
and seven charged states.
Notice nevertheless that the three left handed sneutrinos are 
decoupled 
from the Higgs-right handed sneutrinos,
and also the six charged sleptons 
from the charged Higgses.
The upper bound for the lightest Higgs boson mass turns out to be similar to the one of the NMSSM, about 140 GeV
after imposing the Landau pole constraint up to the GUT scale.

Obviously, the 
{\bf phenomenology} of models where $R$-parity is broken differs substantially
from that of models where $R$-parity is 
conserved. Needless to mention, 
the lightest supersymmetry particle (LSP) 
is no longer stable, and
therefore not all SUSY chains must yield
missing energy events at colliders.
In \cite{Ghosh:2008yh}, the decays of the lightest neutralino
to two body ($W$-lepton) final states were discussed.
The correlations of the decay branching ratios with the neutrino mixing angles were studied as another possible test of the $\mu\nu$SSM at the LHC.
The phenomenology of the $\mu\nu$SSM was also analyzed
in \cite{Hirsch0}, particularized for one and two generations
of right-handed sneutrinos, and taking into account all 
possible final states when studying the decays of the lightest neutralino. 
Possible signatures that might allow to distinguish this model from other R-parity breaking models were discussed qualitatively in these two works \cite{Ghosh:2008yh,Hirsch0}. 


Let us finally discuss potential problems of the $\mu\nu$SSM and their possible solutions.

Since R-parity is broken,
one could add in the superpotential the usual
lepton and baryon number violating terms, $LLe^c + LQd^c$ and $d^cd^cu^c$, 
producing fast {\bf proton decay} (the new terms of the $\mu\nu$SSM are obviously harmless with respect to proton decay).
Nevertheless, the choice of $R$-parity is {\it ad hoc}. There are other discrete 
symmetries, 
like e.g. baryon triality which only forbids the baryon violating operators. 
Obviously, for all these symmetries R-parity is violated.
Besides, in superstring constructions the matter superfields can be located in different sectors of the compact space or have different extra $U(1)$ charges, in such a way that 
some operators violating $R$-parity can be forbidden \cite{old}, 
but others can be allowed. Let us remark here that even if the terms
$LQd^c$ are set to zero at a high-energy scale, one-loop corrections in the
$\mu\nu$SSM will generate them. Nevertheless, these contributions are very small, as shown in \cite{MuNuSSM2}.

In the $\mu\nu$SSM the usual bilinear
$\mu$ term of the MSSM, as well as Majorana masses for neutrinos are absent from the superpotential (\ref{superpotential}), and {\bf only dimensionless trilinear
couplings are present} (the EW scale of the breaking being determined by the
soft terms in the scalar potential).
For this to happen we can invoke a $Z_3$ symmetry 
as it is usually done in the NMSSM. 
Nevetheless,
let us recall that 
this is actually what happens 
in superstring constructions, where the low-energy limit is determined
by the massless superstring modes. Since the massive modes 
have huge masses, of the order of the string scale,
only the trilinear couplings for the massless modes are relevant. 

Since
the superpotential 
has a $Z_3$ symmetry, 
one would expect to have also a 
cosmological {\bf domain wall} problem 
like in the NMSSM. 
Nevertheless, 
the usual solutions to this problem 
will also work in this 
case: non-renormalizable operators 
can break 
explicitly the dangerous $Z_3$ symmetry, lifting the degeneracy of the 
three original vacua, and this can be done without introducing hierarchy 
problems. In addition, these operators can be chosen small enough as 
not to alter the low-energy phenomenology.

When lepton number is broken, {\bf flavour violating processes} are possible.
Although there are strong experimental constraints, these are fulfilled
in the $\mu\nu$SSM once neutrino data are imposed, similar to the
case of BRpV \cite{mario}.

We mentioned above that when R-parity is broken the LSP is not stable.
Thus
the neutralino 
or the sneutrino, 
with very short lifetimes,
are no longer candidates for the {\bf dark matter} (DM) of the Universe.
Nevertheless, other SUSY particles
such as the
gravitino 
or the axino 
can still be used 
since their lifetimes are typically very long. 
Concerning the {\bf gravitino}, it has an interaction term in the supergravity Lagrangian
with 
the photon and the photino. Since
the photino and the left-handed neutrinos are mixed due to the breaking of R-parity,
the gravitino will be able to decay 
into a photon and a neutrino. 
The decay is suppressed both by
the gravitational interaction and by the small R-parity violating coupling, 
thus
its lifetime can be much longer than the age of the 
Universe \cite{Takayama:2000uz}.
Therefore, the gravitino can be a DM candidate.
Since the gravitino decays producing a monochromatic photon with an energy half of the
gravitino mass,
the prospects for detecting these gamma rays
in satellite experiments were analyzed in the context of 
bilinear R-parity violation scenarios in the literature.
In a recent work \cite{recentgravitino}, gravitino 
DM and
its possible detection in the FERMI satellite were discussed in the context of the $\mu\nu$SSM.
Gravitino masses larger than $20$ GeV are disfavored by 
the 
diffuse photon background measurements, but
a gravitino with a mass range between $0.1 - 20$ GeV gives rise to 
a signal that might be observed by the FERMI satellite.

\bibliographystyle{aipproc}   


\IfFileExists{\jobname.bbl}{}
 {\typeout{}
  \typeout{******************************************}
  \typeout{** Please run "bibtex \jobname" to optain}
  \typeout{** the bibliography and then re-run LaTeX}
  \typeout{** twice to fix the references!}
  \typeout{******************************************}
  \typeout{}
 }


\end{document}